\documentstyle[aps,psfig,multicol]{revtex}
\begin{document} 
\draft
\title{Spectral function of transverse spin fluctuations
in an antiferromagnet}
\author{Avinash Singh}
\address{Department of Physics, Indian Institute of Technology, Kanpur-208016, India}
\maketitle
\begin{abstract} 
The spectral function of transverse spin fluctuations,
including the contributions from both the 
single-particle and the collective (magnon) excitations
in an antiferromagnet, is evaluated for the Hubbard model with
NN and NNN hoppings in the full $U$-range from weak coupling
to strong coupling. For the NN hopping model,
the magnon excitations are dominant for $U>2.5$ ($d=2$),
so that an effective spin description of the AF state holds
down to a surprisingly low $U$ value.
In the weak coupling limit the spectral function is suppressed 
at low energies, as if due to an effective gap. 
NNN hopping $t'$ leads to magnon softening
and also a significant increase in the low-energy spectral function
due to single-particle excitations. 
Evolution of the magnon spectrum with $t'$ is studied
in the strong coupling limit,
and the quantum spin-fluctuation correction to
sublattice magnetization in $d=2$ and the N\'{e}el temperature
in $d=3$ are also evaluated.
\end{abstract}
\pacs{75.10.Jm, 75.10.Lp, 75.30.Ds, 75.10.Hk}  

\begin{multicols}{2}\narrowtext
\section{Introduction}
The antiferromagnetic state of the half-filled Hubbard model 
with nearest-neighbour (NN) hopping
is characterized by an energy gap, and 
the spectrum of transverse spin fluctuations consists of the low-lying,
collective (magnon) excitations, as well as the single-particle
excitations across the gap. 
Generally, there is a clear distinction between these two
excitations which are well separated in energy.
However, recent extensions, e.g. with 
disorder,\cite{dis} impurities,\cite{denteneer} and next-nearest-neighbour (NNN)
hopping,\cite{nnn} clearly show the existence of essentially gapless
antiferromagnetism, arising from a variety of mechanisms.
Thus, with increasing disorder (on-site potential disorder)
the two Hubbard bands progressively broaden, until the band gap vanishes
when the disorder strength $W \approx U$.
With low-$U$ impurities on the other hand,
the effective charge gap becomes negligible
due to nearly localized states on the impurities.\cite{denteneer}
And lastly, when NNN hopping is included, 
a band asymmetry is introduced which
reduces the band gap in the AF state, and in weak coupling there exists
a region of gapless AF phase in the magnetic phase diagram.\cite{nnn}

This gaplessness implies that the collective magnon excitations are
no longer well defined, and actually merge with the single-particle
excitations, thereby necessitating a unified scheme for the
evaluation of transverse spin fluctuation.
While the magnon contribution to the transverse spin
fluctuations was evaluated recently, and the sublattice magnetization
and the N\'{e}el temperature were obtained within a
renormalized spin fluctuation theory
in the whole $U/t$ range,\cite{trans,neel}
in this paper we describe an alternate scheme for
evaluating the transverse spin fluctuations which
allows both the collective excitations and the single-particle
excitations to be studied on the same footing.

Single-particle excitations are especially significant because 
it is precisely this part of the transverse spin spectral
function which allows for a quantitative distinction of the 
antiferromagnetic state within the Hubbard model 
from that of an equivalent Heisenberg spin model
with $U$-dependent, extended-range spin couplings $J_{ij}(U)$,
but possessing only magnon excitations.
A study of the relative strengths of the magnon and single-particle
excitations, in terms of the integrated spectral weights in
the whole $U/t$ range,
will therefore allow for a quantitative demarkation along the $U/t$ axis
below which single-particle excitations are the dominant contribution.
Thus, while in this low-$U$ regime the AF state is not well
described in terms of an effective Heisenberg spin model possessing
only magnon excitations, in the intermediate and strong coupling limits
use of an effective Heisenberg model with $U$-dependent spin couplings
$J_{ij}(U)$,\cite{logan.book} and generally use of a
spin picture\cite{trans,neel} is appropriate.

Yet another significance of the single-particle excitations
is interestingly connected with the spin commutation relation
$[S^+, S^-]=2S^z$. The RPA-level ground-state expectation value 
$\langle [S^+ , S^- ]\rangle_{\rm RPA}$,
involving the difference of transverse spin correlations 
%$\langle S^+ S^- \rangle_{\rm RPA}$ and  
%$\langle S^- S^+ \rangle_{\rm RPA}$
evaluated in the AF ground state,
should be identically equal to $\langle 2S^z \rangle_{\rm HF}$,
the local magnetization at the HF level.
This is deduced from the fact that both RPA and HF approximations are
of the same order (O(1)) within the
systematic inverse-degeneracy expansion scheme\cite{quantum}
in powers of $1/{\cal N}$, where ${\cal N}$ is the number of orbitals per site.
Therefore both RPA and HF become exact in the limit ${\cal N} \rightarrow\infty$,
when all corrections of order $1/{\cal N}$ or higher vanish.
The magnon contribution 
$\langle [S^+ , S^- ] \rangle_{\rm RPA} ^{\rm magnon}$
was indeed found to be in excellent agreement with the HF magnetization
$\langle 2S^z \rangle_{\rm HF}$
in the intermediate and strong coupling limits.\cite{trans}
However, a discrepancy was observed at small $U$ 
which was attributed to the neglect of the single-particle excitations, 
which become relatively more important in the weak coupling limit.
We will show here that indeed when the
single-particle excitations are included this discrepancy is exactly removed.

Finally, the unified scheme for incorporating both the 
single particle and collective excitations in 
the evaluation yields the complete
spectrum of transverse spin excitations in the magnetic 
state. Calculations with realistic models of magnetic
solids can hence be used for comparison with
results of scattering studies with neutrons, which
are direct experimental probes into  
the spectrum of magnetic excitations in solids.
In this regard the necessity of more realistic models
which include NNN hopping etc. has been acknowledged recently
from realistic band structure studies, photoemission data
and neutron-scattering measurements of high-T$_{\rm c}$
and related materials.\cite{nnn1,nnn2,nnn3,nnn4}

\section{Hubbard model with next-nearest-neighbour hopping}
We consider the following Hamiltonian on a square lattice,
with hopping terms $t$ and $t'$ between 
nearest-neighbour (NN) and next-nearest-neighbour (NNN) 
pairs of sites $\langle ij\rangle$ and $\langle ik\rangle$
respectively, 
\begin{equation}
H = 
-t \sum_{\langle {ij} \rangle \sigma} ^{\rm NN}
a_{i \sigma}^{\dagger} a_{j \sigma}
-t' \sum_{\langle {ik} \rangle \sigma} ^{\rm NNN} 
a_{i \sigma}^{\dagger} a_{k \sigma}
+  U\sum_{i} n_{i \uparrow} n_{i \downarrow} \; .
\end{equation}
Extension to the simple cubic lattice is considered in the Appendix.
In the plane-wave
basis defined by $a_{i\sigma}=\sqrt{\frac{1}{N}} \sum_{\bf k} 
e^{i{\bf k}.{\bf r}_i} a_{{\bf k}\sigma}$,
the non-interacting part of the Hamiltonian 
$H_0=\sum_{{\bf k}\sigma} (\epsilon_{\bf k} + 
\epsilon' _{\bf k} )a_{{\bf k}\sigma} ^\dagger
a_{{\bf k}\sigma} $, where $\epsilon_{\bf k}$ and 
$\epsilon'_{\bf k}$ 
are the two free-particle energies,
correponding to NN and NNN hopping respectively, 
\begin{eqnarray}
\epsilon_{\bf k} &=& -2t(\cos k_x + \cos k_y ) \nonumber \\
\epsilon'_{\bf k}&=& -4t' \cos k_x. \cos k_y  .
\end{eqnarray}

For the NN model,
the HF-level description of the broken-symmetry AF state,
and the transverse spin fluctuations
have been discussed earlier in the strong,\cite{spfluc}
intermediate,\cite{inter,weak} and weak coupling\cite{weak} limits.
We briefly discuss the extension for the NNN hopping.
Since the NNN hopping term connects sites in the same sublattice,
in the two-sublattice basis the  $\epsilon'_{\bf k}$ term
appears in the diagonal matrix elements
of the HF Hamiltonian
\begin{equation}
H_{\rm HF}^\sigma ({\bf k})=
\left [ \begin{array}{cc}
-\sigma \Delta + \epsilon'_{\bf k} & \epsilon_{\bf k}  \\
\epsilon_{\bf k} & \sigma \Delta+ \epsilon'_{\bf k}   \end{array} \right ]
= \epsilon'_{\bf k} \; {\bf 1} + 
\left [ \begin{array}{cc}
-\sigma \Delta & \epsilon_{\bf k}  \\
\epsilon_{\bf k} & \sigma \Delta  \end{array} \right ]
\end{equation}
for spin $\sigma$. Here $2\Delta=mU$,
where $m$ is the sublattice magnetization.
For the NN hopping model $2\Delta$ is also the energy gap
for single-particle excitations.
Since the $\epsilon' _{\bf k}$ term appears as a unit matrix,
the eigenvectors of the HF Hamiltonian remain unchanged
from the NN case,\cite{spfluc} 
whereas the eigenvalues correponding to the quasiparticle
energies are modified to,
\begin{equation}
E_{{\bf k}\sigma}^{(\pm)}=\epsilon'_{\bf k} 
\pm \sqrt{\Delta^2 + \epsilon_{\bf k} ^2}.
\end{equation}
The two signs $\pm$ refer to the two quasiparticle bands.
The band gap is thus affected  by the NNN hopping term,
and it progressively decreases 
as $2\Delta - 4t'$ in the weak coupling limit. 

As the eigenvectors of the HF Hamiltonian are  unchanged,
the self-consistency condition retains its form,
and therefore the sublattice magnetization is independent of $t'$,
provided there is a nonzero band gap, 
with the lower band occupied and the upper band empty.
We will restrict ourselves only to this regime where there is
no band overlap, though the gap may vanish when 
the two bands are just touching.  
The fermionic quasiparticle amplitudes
$a_{{\bf k}\sigma}$ and $b_{{\bf k}\sigma}$ 
for spin $\sigma=\uparrow,\downarrow$
and the two quasiparticle bands are given by
\begin{eqnarray}
a_{{\bf k}\uparrow\ominus}^2 =
b_{{\bf k}\downarrow\ominus}^2 =
a_{{\bf k}\downarrow\oplus}^2 =
b_{{\bf k}\uparrow\oplus}^2 &=& \frac{1}{2}\left 
( 1+ \frac{\Delta}{\sqrt{\Delta^2 +\epsilon_{\bf k} ^2}} \right ) \nonumber \\
a_{{\bf k}\uparrow\oplus}^2 =
b_{{\bf k}\downarrow\oplus}^2 =
a_{{\bf k}\downarrow\ominus}^2 =
b_{{\bf k}\uparrow\ominus}^2 &=& \frac{1}{2}\left 
( 1- \frac{\Delta}{\sqrt{\Delta^2 +\epsilon_{\bf k} ^2}} \right ).
\end{eqnarray}
These relationships follow from the spin-sublattice and particle-hole 
symmetry in the AF state of the half-filled system.

However, this situation changes when with increasing
NNN hopping $t'$ the two bands overlap and the band gap vanishes. 
This overlap indicates that the low-lying states 
in the upper band corresponding to double occupancy 
are lower in energy than the high-energy states in the lower band.
This results in charge transfer and consequently 
partially empty and doubly occupied sites. 
The situation is therefore analogous to the
addition of holes or electrons in the half-filled
Hubbard model, with all its associated complications
of spin bags, strings of upturned spins,
spiral and striped phases etc.\cite{kampf+brenig} 
Thus the simple gapless AF state of the NNN hopping model,\cite{nnn}
which is indeed a self-consistent HF solution,
may actually be unstable due to same instabilities as the hole-doped AF.
Numerical HF studies do show this instability,
and will be further discussed elsewhere.\cite{tbp}

\section{The transverse spin spectral function}
The spectral function for transverse spin fluctuations
in the AF state is obtained from the imaginary part 
of the corresponding time-ordered propagator 
of the transverse spin operators $S_i ^-$ and $S_j ^+$ at 
sites $i$ and $j$,
$\chi^{-+}({\bf q}\omega)=\int dt \sum_i 
e^{i\omega(t-t')} e^{-i{\bf q}.({\bf r}_i -{\bf r}_j)} 
\langle \Psi_{\rm G} | T [ S_i ^- (t) S_j ^+ (t')]|\Psi_{\rm G}\rangle$.
In terms of the RPA expression $[\chi^{-+}({\bf q}\omega)]_{\rm RPA}=
[\chi^0 ({\bf q}\omega)]/[1-U\chi^0 ({\bf q}\omega)]$,
where $\chi^0 ({\bf q}\omega)$ is the zeroth-order particle-hole
propagator, 
there are two sources of contribution to the spectral function, 
$A(\omega)=\sum_{\bf q} {\rm Im}\;{\rm Tr}[\chi^{-+}({\bf q}\omega)]$.
The contribution arising from the imaginary part of 
$\chi^0 ({\bf q}\omega)$ is
associated with the single-particle excitations across the gap
(with $\omega>2\Delta$ for the NN model), 
whereas that due to the vanishing of the
denominator $1-U\chi^0 ({\bf q}\omega)=0$ is  
associated with the collective magnon excitations
(involving $\omega<2\Delta$ for the NN model).
As mentioned earlier,
this latter contribution to the integrated spectral weight
$\pi ^{-1}\int d\omega A(\omega)$, and therefore
to the transverse spin correlations 
$\langle S^+ S^- \rangle_{\rm RPA}$ and $\langle S^- S^+ \rangle_{\rm RPA}$ 
was recently studied in the context of
quantum correction to sublattice magnetization and the N\'{e}el
temperature in the whole $U/t$ range.\cite{trans,neel}

The evaluation of the 
transverse spin spectral function, including contributions from both 
the single-particle and collective excitations,
is facilitated by expressing 
the $2\times 2$ complex matrix $[\chi^0({\bf q}\omega)]$
in terms of its eigenvalues  
$\lambda_{\bf q} ^n (\omega)$ and 
eigenvectors $|\phi_{\bf q} ^n (\omega)\rangle $, and we obtain 
\begin{eqnarray}
A(\omega) &=& \sum_{\bf q} {\rm Im} {\rm Tr} 
[\chi^{-+}({\bf q}\omega)]
= \sum_{\bf q} {\rm Im} {\rm Tr} 
\frac{[\chi^0 ({\bf q}\omega)]}{{\bf 1}-U[\chi^0 ({\bf q}\omega)]}
\nonumber \\
&=& \sum_{\bf q} {\rm Im} {\rm Tr} \sum_{n=1,2} \left 
( \frac{\lambda_{\bf q} ^n (\omega)}
{1-U\lambda_{\bf q} ^n (\omega)} \right ) |\phi_{\bf q} ^n (\omega)\rangle 
\langle \phi_{\bf q} ^n (\omega) | \nonumber \\
&=& \sum_{\bf q} {\rm Im}  \sum_{n=1,2} \left 
( \frac{\lambda_{\bf q} ^n (\omega)}
{1-U\lambda_{\bf q} ^n (\omega)} \right ).
\end{eqnarray}

If instead of taking the trace in Eq. (6),
the two diagonal matrix elements are considered separately,
then we obtain the transverse spin correlations on the A(B) sublattice 
by integrating over frequency:
\begin{eqnarray}
\langle S^+ S^- \rangle _{\rm B(A)} = 
\langle S^- S^+ \rangle _{\rm A(B)} = 
\int \frac{d\omega}{\pi}
\sum_{\bf q} {\rm Im}  [\chi^{-+}({\bf q}\omega)]_{\rm A(B)} \nonumber \\
= \int \frac{d\omega}{\pi}
\sum_{\bf q} \sum_{n=1,2} {\rm Im}\left ( \frac{\lambda_{\bf q} ^n (\omega)}
{1-U\lambda_{\bf q} ^n (\omega)}\right ) 
|\phi_{\bf q} ^n (\omega) |^2 _{\rm A(B)}  .
\end{eqnarray}
Here the relationship 
$\langle S^+ S^- \rangle _{\rm B(A)} = 
\langle S^- S^+ \rangle _{\rm A(B)}$ between the transverse spin
correlations on opposite sublattices follows from the 
spin-sublattice symmetry in the AF state. 

\section{Computational procedure and Results}
To begin with the zeroth order
$2\times 2$ matrix $[\chi^0({\bf q}\omega)]= 
i\int \frac{d\omega}{2\pi} \sum_{\bf k}'
[G^\uparrow ({\bf k}'\omega')][G^\downarrow ({\bf k'-q},\omega'-\omega)]$, 
is evaluated in the broken-symmetry AF state. 
In terms of the fermionic quasiparticle amplitudes and energies,
$[\chi^0({\bf q}\omega)]$ is given by,\cite{spfluc}
\end{multicols}

\widetext
\begin{eqnarray}
[\chi^0({\bf q}\omega)] &=& \sum_{\bf k}
\left [ \begin{array}{lr} 
a_{{\bf k}\uparrow \ominus}^2  a_{{\bf k-q}\downarrow \oplus}^2  & 
a_{{\bf k}\uparrow \ominus}b_{{\bf k}\uparrow \ominus}a_{{\bf k-q}\downarrow \oplus}b_{{\bf k-q}\downarrow \oplus} \\
a_{{\bf k}\uparrow \ominus}b_{{\bf k}\uparrow \ominus}a_{{\bf k-q}\downarrow \oplus}b_{{\bf k-q}\downarrow \oplus} &
b_{{\bf k}\uparrow \ominus}^2  b_{{\bf k-q}\downarrow \oplus}^2  \end{array}
\right ] 
\frac{1}{E_{{\bf k-q}}^{\oplus} - E_{\bf k} ^{\ominus} + \omega -i \eta}
\nonumber \\
&+& \sum_{\bf k}
\left [ \begin{array}{lr} 
a_{{\bf k}\uparrow \oplus}^2  a_{{\bf k-q}\downarrow \ominus}^2  & 
a_{{\bf k}\uparrow \oplus}b_{{\bf k}\uparrow \oplus}a_{{\bf k-q}\downarrow \ominus}b_{{\bf k-q}\downarrow \ominus} \\
a_{{\bf k}\uparrow \oplus}b_{{\bf k}\uparrow \oplus}a_{{\bf k-q}\downarrow \ominus}b_{{\bf k-q}\downarrow \ominus} &
b_{{\bf k}\uparrow \oplus}^2  b_{{\bf k-q}\downarrow \ominus}^2  \end{array}
\right ] 
\frac{1}{E_{{\bf k}}^{\oplus} - E_{{\bf k-q}} ^{\ominus} - \omega -i \eta} .
\end{eqnarray}

\begin{multicols}{2}\narrowtext
\noindent
Analytical evaluation of $[\chi^0 ({\bf q}\omega)]$
in the strong and intermediate coupling limits has been discussed
earlier for the NN hopping model,\cite{spfluc,inter}
and is described in the Appendix for the NNN hopping model.
For arbitrary $U$, the ${\bf k}$-sum is numerically performed 
to evaluate the complex matrix $[\chi^0 ({\bf q}\omega)]$,
which is then diagonalized to obtain the two eigenvalues
$\lambda_{\bf q} ^n$ and eigenvectors $|\phi_{\bf q} ^n \rangle$.
The infinitesimal $\eta$ is appropriately chosen according to the
fineness of the ${\bf k}$ and ${\bf q}$ grids.
Typically, the grid sizes $dk$ and $dq$ were taken of order 0.05
and $\eta \sim 0.01$. 
Summing over the whole range of ${\bf q}$ values 
between $0$ and $\pi$ in each
dimension then yields the spectral function $A(\omega)$
from Eq. (6),
and the transverse spin correlations
from Eq. (7).

The integrated spectral weight is obtained by numerically
integrating the spectral function $A(\omega)$ over frequency.
The collective and single-particle contributions can be
evaluated separately by integrating over the $\omega$-regions
$\omega<2\Delta$ and $\omega>2\Delta$  respectively
(for the NN case).
As the single-particle excitations have a continuum distribution
in Eq. (6), the evaluation of this contribution
is less computationally intensive.
However, for collective excitations, which are a set of delta functions, 
it is necessary to have in the numerical integration procedure
fine enough $\omega$- and $q$-grids, 
so that a sufficiently large number of 
magnon modes are picked up in the $\omega$ and $q$ sums.
The method described in ref. \cite{trans,neel} is more
efficient for evaluating the  collective-excitation contribution.

Before discussing the results we first consider the two limiting cases.
In the limit of vanishing interaction strength 
$U\rightarrow 0$, $\Delta\rightarrow 0$, we have $\chi^{-+}({\bf q}\omega)
=\chi^0 ({\bf q}\omega)$, and therefore the magnon contribution vanishes. 
Also, as all the quasiparticle probabilities $a^2 = b^2 = 1/2$
from Eq. (5), we have from above
\begin{eqnarray}
& & {\rm Im} \chi^0({\bf q}\omega)
= \sum_{\bf k} \left [ \begin{array}{lr} 
\frac{1}{4}\;\;\;  &  \frac{1}{4}\\
\frac{1}{4}\;\;\;  & \frac{1}{4} \end{array}
\right ] \times \pi \times \nonumber \\
& &  [ \delta(E_{\bf k-q}^\oplus - E_{\bf k}^\ominus +\omega) 
+ \delta(E_{\bf k}^\oplus + E_{\bf k-q}^\ominus - \omega) ].
\end{eqnarray}
In the non-interacting limit, the integrated spectral 
weight therefore yields  
\begin{equation}
\stackrel{\rm Lim}{U\rightarrow 0} \;
\int \frac{d\omega}{\pi}
\sum_{\bf q} {\rm Im} {\rm Tr} [\chi^{-+}({\bf q}\omega)] = 1/2 .
\end{equation}

On the other hand, in the strong coupling limit 
$U\rightarrow \infty$, the quasiparticle bands
are narrowed to infinitesimal width, so that
the imaginary part of $\chi^0({\bf q}\omega)$ strengthens to
delta functions at the gap frequencies $\omega=\pm U$.

\begin{figure}
\vspace*{-60mm}
\hspace*{-28mm}
\psfig{file=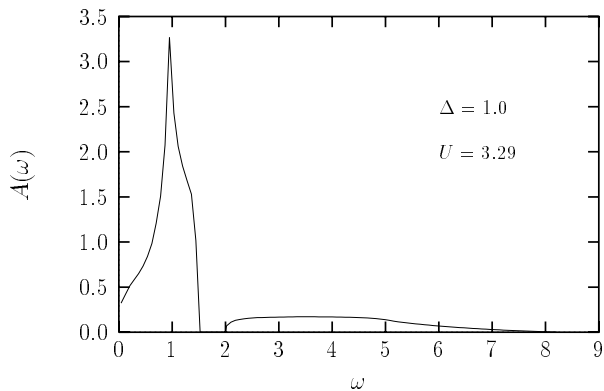,width=135mm,angle=0}
\vspace{-70mm}
\caption{
The spectral function for transverse spin fluctuations obtained from Eq. (6),
showing a distinct separation between the low-lying, collective excitations 
$(\omega<2\Delta)$ and the single-particle excitations $(\omega>2\Delta)$.}
\end{figure}

\begin{figure}
\vspace*{-60mm}
\hspace*{-28mm}
\psfig{file=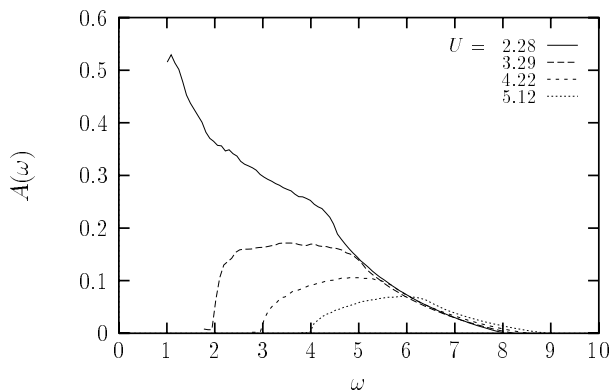,width=135mm,angle=0}
\vspace{-70mm}
\caption{The single-particle contribution to the transverse-spin
spectral function $A(\omega)=\sum_q {\rm Im} \; {\rm Tr} \chi^{-+}(q\omega)$,
for gap values $2\Delta=1,2,3,4$, showing the rapid decrease in the spectral weight
with increasing correlation.}
\end{figure}

\noindent
Therefore, for both possible values of $\chi^0({\bf q}\omega)$,
zero or infinity, the single-particle contribution
to the imaginary part of
$\chi^{-+}({\bf q}\omega)$ vanishes.
Thus, with increasing interaction strength,
the spectrum of transverse spin fluctuations changes from 
predominantly single-particle excitations in weak coupling
to predominantly magnon excitations in the strong coupling limit.

Figures 1 through 6 depict results for the NN  model.
Fig. 1 shows a typical transverse spin spectral function
$A(\omega)=\sum_q {\rm Im} \; {\rm Tr} \chi^{-+}({\bf q}\omega)$,
for a moderate correlation value, showing the 
distinct separation between the low-lying, collective excitations 
$(\omega<2\Delta)$ and the single-particle excitations $(\omega>2\Delta)$.

\begin{figure}
\vspace*{-60mm}
\hspace*{-28mm}
\psfig{file=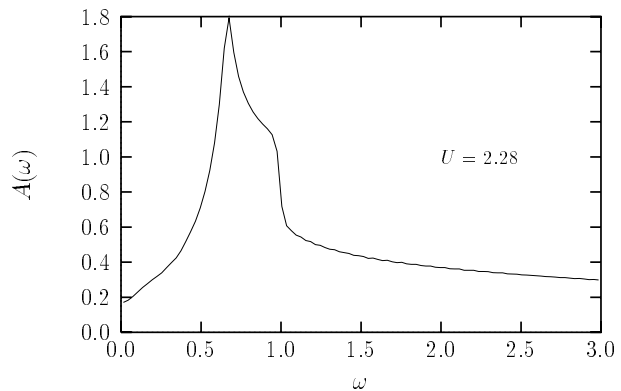,width=135mm,angle=0}
\vspace{-70mm}
\caption{
The near merging of the collective $(\omega<2\Delta)$ and the single-particle
$(\omega>2\Delta)$ contributions to the transverse-spin spectral function
for $2\Delta =1$.}
\end{figure}

\begin{figure}
\vspace*{-65mm}
\hspace*{-28mm}
\psfig{file=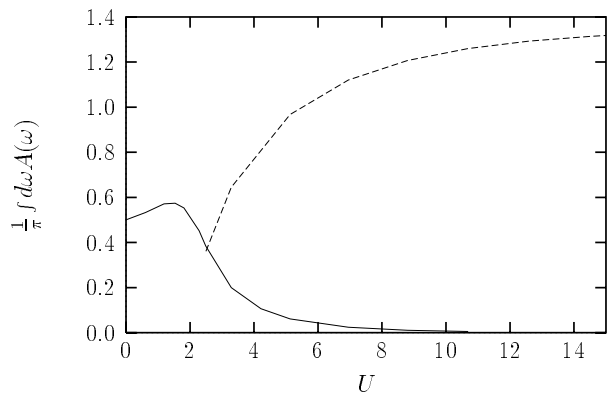,width=135mm,angle=0}
\vspace{-70mm}
\caption{
The integrated spectral weights for the single-particle (solid)
and collective (dashed) excitations, 
showing their strong suppression respectively 
in the strong and weak coupling limits.
$U\approx 2.5$ marks the point below
which an effective spin description of the antiferromagnetic state
within the Hubbard model starts to fail.}  
\end{figure}

Fig. 2 shows the single-particle contribution to the 
spectral function for moderate gap values $2\Delta=1,2,3,4$.
The rapid decrease in the spectral weight with increasing correlation
shows that the single-particle excitations are strongly
suppressed in the strong correlation limit. 

With decreasing correlation, the separation between the low-lying
collective excitations and the single-particle excitations progressively
decreases. Fig. 3 shows the near merging of the collective $(\omega<2\Delta)$ and the single-particle
$(\omega>2\Delta)$ excitations for $2\Delta =1$ ($U=2.28$).

A comparison of the integrated spectral weights 
as a function of $U$ is shown in Fig. 4, 
for the single-particle and collective excitations.
While the collective excitations are seen to 
sharply fall-off in the weak coupling limit,

%\newpage
\begin{figure}
\vspace*{-65mm}
\hspace*{-28mm}
\psfig{file=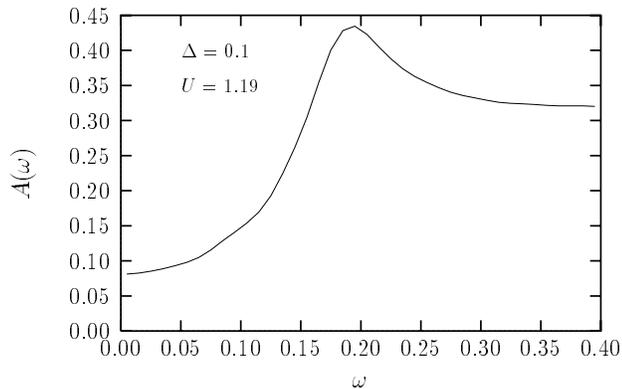,width=135mm,angle=0}
\vspace{-70mm}
\caption{
The low-energy part of the 
transverse-spin spectral function for very low $\Delta$,
showing a gap-like suppression of excitations at low energies.} 
\end{figure}

\begin{figure}
\vspace*{-70mm}
\hspace*{-28mm}
\psfig{file=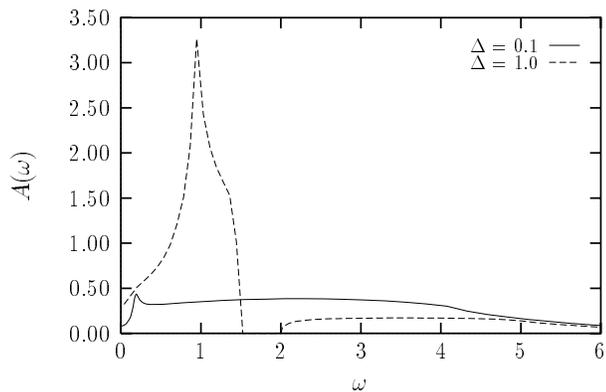,width=135mm,angle=0}
\vspace{-70mm}
\caption{
The drastic suppression in the contribution of
the low-energy, collective excitations in going from 
$U=3.29$ ($\Delta=1.0$) to $U=1.19$ $(\Delta=0.1)$.}
\end{figure}

\noindent
the single-particle
excitations are likewise 
suppressed in the strong coupling limit.
For $U > 2.5$, the collective (magnon) excitations are dominant,
and thus an effective spin description of the antiferromagnetic
state is appropriate down to a surprisingly low $U$ value.

We now examine the role of the single-particle excitations in
the sum rule following from the spin commutation relation
$[S^+, S^-]=2S^z$. As discussed in the Introduction,
the RPA-level, AF ground-state expectation value
$\langle [S^+ _i , S^- _i ]\rangle_{\rm RPA}$,
including both magnon and single-particle contributions,
should be identically equal to $\langle 2S^z _i \rangle_{\rm HF}$.
The single-particle contribution to this expectation value,
obtained from the transverse spin correlations
$\langle S^+  S^-  \rangle$ and $\langle S^-  S^+  \rangle$
evaluated from Eq. (7),
are shown in Table I for several $U$ values. 
Also shown are the magnon contributions which were obtained
earlier.\cite{trans,neel}
The total of the magnon and single-particle contributions
is indeed in close agreement with the 
HF magnetization $\langle 2S^z _i \rangle_{\rm HF}$.

\begin{figure}
\vspace*{-60mm}
\hspace*{-28mm}
\psfig{file=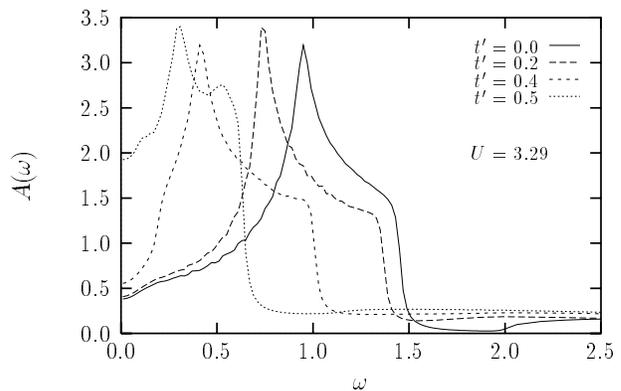,width=135mm,angle=0}
\vspace{-70mm}
\caption{
Effects of the NNN hopping $t'$ on the 
transverse spin spectral function, 
showing  the shift in the magnon spectrum towards lower energy
with increasing $t'$.}
\end{figure}

The sharp fall-off of the gapless magnon contribution 
in the weak coupling limit 
essentially leaves only the single-particle contribution.
As these excitations have a minimum-energy threshold of $2\Delta$,
the spectrum of trans-
verse spin fluctuations essentially acquires
a pseudo gap at low energies. 
This is clearly seen in Fig. 5,
showing a gap-like suppression in the density of excitations
at low energies.

Fig. 6 shows 
a comparison of the spectral functions for $\Delta=0.1$ 
and $\Delta=1.0$. The drastic suppression in the contribution of
the low-energy, collective excitations in going from 
$U=3.29$ ($\Delta=1.0$) to $U=1.19$ $(\Delta=0.1)$
is quite remarkable.

Fig. 7 shows the effects of the NNN hopping on the
spectral function. With increasing NNN hopping $t'$,
the magnon spectrum shifts towards lower energy, 
which reflects the magnon softening.
There is also a substantial increase in the low-energy
spectral function due to the reduction in the energy gap
with NNN hopping.
For $U=3.29$, we have $\Delta=1$,
and the energy gap $2\Delta - 4t'$ just vanishes when $t'=0.5$.
It is therefore clear that the strong enhancement in the
low-energy spectral function for $t'=0.5$ is due to the
gapless single-particle excitations.

\begin{table}
\caption{The RPA-level expectation values of the spin commutator 
$[S^+ _i,S^- _i]$, 
evaluated in the AF state from the transverse spin propagator,
with contributions from the collective (magnon) excitations, 
and the single-particle excitations.
The total is compared with the HF magnetization 
$\langle 2S^z _i\rangle_{\rm HF}$.}
\begin{tabular}{cccccc}
$\Delta$ & $U$ & $\langle[S^+ _i,S^- _i ]\rangle_{\rm mag}$ & 
$\langle[S^+ _i,S^- _i]\rangle_{\rm sp}$ & $\langle[S^+ _i,S^- _i]
\rangle_{\rm tot}$ & $\langle 2S^z _i\rangle_{\rm HF}$ \\ \tableline
0.6 & 2.499 &  .315 &  .172 &  .487 & .480 \\
1.0 & 3.292 &  .553 &  .064 &  .617 & .607  \\
2.0 & 5.127 &  .776 &  .009 &  .785 & .781   \\
3.0 & 6.948 &  .865 &       &       & .863   \\
\end{tabular}
\end{table}

\newpage
This magnon softening has been analytically studied
in the strong coupling limit, and the results are given
in the Appendix. Due to a frustration induced by the
NNN hopping term, the long-wavelength magnon-mode energy
is suppressed, and for the simple cubic lattice 
it vanishes at $t'=1/2$.
At and above this critical value the N\'{e}el temperature
$T_{\rm N}$ vanishes, indicating that the N\'{e}el state
is unstable. The new phase which
is stabilized beyond this point is a F-AF phase,
involving antiferromagnetic ordering 
of spins in planes (say parallel to the x-y plane),
and ferromagnetic alignment of spins along the z direction.\cite{tbp}
Hence $t'=1/2$ marks the phase boundary
between the AF phase with ordering wavevector
${\bf Q}=(\pi,\pi,\pi)$ and the F-AF phase with
${\bf Q}=(\pi,\pi,0)$.

For large but finite $U$, there is a relative stiffening of the magnon
modes due to the 3rd neighbour ferromagnetic coupling induced
by the NNN hopping. This reflects a reduction in the degree of
frustration, so that a slightly higher $t'$ value is required to
suppress the magnon velocity to zero. From a study of the $U$-dependence
of this critical $t'$ value, the magnetic phase diagram of the three
dimensional Hubbard model with NNN hopping has been obtained
recently.\cite{tbp}

\section{Conclusions}
In conclusion, we have studied the spectral function
of transverse spin fluctuations in an antiferromagnet
using a unified approach
which includes the contributions from both 
single-particle excitations and collective magnon excitations.
For the NN hopping model in two dimensions,
the integrated spectral weight was studied
in the whole $U$ range, and the collective magnon contribution
was found to be dominant for $U>2.5$,
so that an effective spin description of the AF state is
appropriate down to a surprisingly low $U$ value.
In the weak coupling limit 
the sharp fall-off of the gapless magnon contribution
essentially leaves only the single-particle contribution
having a minimum-energy threshold of $2\Delta$,
so that the spectrum of transverse spin fluctuations effectively
acquires a gap at low energies. 
The evolution of the spectral function with
increasing NNN hopping, which reduces the energy gap,
shows a shift of the spectral function
towards lower energy due to magnon softening,
and also a significant rise at low energy due to 
single-particle excitations. 

\end{multicols}
\widetext
\newpage
\section*{Appendix}
\subsection*{Hubbard model with NNN hopping --- strong coupling limit}
In this section we describe for completeness the
AF properties of the half-filled Hubbard model with NNN hopping in the
strong coupling limit. NNN hopping introduces a frustration 
in the system through the competing NNN AF interaction which
enhances spin fluctuations and destabilizes the N\'e{e}l state.
Focussing on the enhancement of transverse spin
fluctuations at the RPA level by NNN hopping,
we examine the spectrum of the collective
(magnon) excitations, and their contribution to the 
quantum spin-fluctuation correction $\delta m_{\rm SF}$
to the sublattice magnetization
in two dimensions, and the reduction in the N\'{e}el temperature
$T_{\rm N}$ in three dimensions. 

As discussed earlier, the quasiparticle amplitudes 
$a_{{\bf k}\sigma(\pm)}$ and $b_{{\bf k}\sigma(\pm)}$ 
which form the eigenvectors of the HF Hamiltonian
remain unchanged by  the NNN hopping $t'$,
provided the charge gap is finite, which is very much so in
the strong coupling limit.
Therefore the only change in $\chi^0 ({\bf q}\omega)$ arises from the
change in the quasiparticle energy expression given in Eq. (4),
which appear in the energy denominators in Eq. (8), and we obtain, 
for the AA matrix element, for example

\begin{eqnarray}
[\chi^0 ({\bf q}\omega)]_{\rm AA} &=&
\sum_{\bf k} \frac{
a_{{\bf k}\uparrow \ominus}^2  a_{{\bf k-q}\downarrow \oplus}^2   }
{\sqrt{\Delta^2 + \epsilon_{\bf k} ^2 } +
\sqrt{\Delta^2 + \epsilon_{{\bf k-q}} ^2} + 
(\epsilon'_{\bf k-q} - \epsilon'_{\bf k} ) + \omega } \nonumber \\
&+&
\sum_{\bf k} \frac{
a_{{\bf k}\uparrow \oplus}^2  a_{{\bf k-q}\downarrow \ominus}^2  }
{\sqrt{\Delta^2 + \epsilon_{\bf k} ^2 } +
\sqrt{\Delta^2 + \epsilon_{{\bf k-q}} ^2} + 
(\epsilon'_{\bf k} - \epsilon'_{\bf k-q} ) - \omega }  \; .
\end{eqnarray}
Substituting 
$a_{{\bf k}\uparrow\ominus}^2 = a_{{\bf k}\downarrow\oplus}^2
\approx  1-\epsilon_{\bf k} ^2/4\Delta^2$ and
$a_{{\bf k}\uparrow\oplus}^2 = a_{{\bf k}\downarrow\ominus}^2
\approx  \epsilon_{\bf k} ^2/4\Delta^2$
in the strong coupling limit,
expanding the denominator in powers of
$t/\Delta$, $t'/\Delta$, $\omega/\Delta$, and 
systematically retaining terms only up to
order $t^2/\Delta^2$ and $t^{'2}/\Delta^2$,
we obtain after performing the ${\bf k}$-sums in two dimensions
for a square lattice with 
$\sum_{\bf k} \epsilon_{\bf k} ^2 = 4t^2$,
$\sum_{\bf k} \epsilon_{\bf k} ^{'2} = 4t^{'2}$,
$\sum_{\bf k} \epsilon'_{\bf k} \epsilon'_{\bf k-q} = 
4t^{'2}\cos q_x \cos q_y$,
\begin{eqnarray}
[\chi^0 ({\bf q}\omega)]_{\rm AA} &=&
\frac{1}{2\Delta} \left [
1-\frac{4t^2}{\Delta^2} + \frac{2t^{'2}}{\Delta^2}(1-\cos q_x \cos q_y)
-\frac{\omega}{2\Delta} \right ] \nonumber \\
&=& 
\frac{1}{U} \left [
1-\frac{2t^2}{\Delta^2}\left(1+\frac{\omega}{2J}\right )
 + \frac{2t^{'2}}{\Delta^2}(1-\cos q_x \cos q_y) \right ],
\end{eqnarray}
where $2\Delta=mU \approx (1-2t^2/\Delta^2)U$
and $J=4t^2/U$.
Similarly evaluating the other matrix elements,\cite{spfluc},
we obtain
\begin{equation}
[1-U\chi^0 ({\bf q}\omega)] =
\frac{2t^2}{\Delta^2} \left [
\begin{array}{cc}
1 - \frac{J'}{J}(1-\gamma ' _{\bf q})+\frac{\omega}{2J} & \gamma_{\bf q} \\
\gamma_{\bf q} & 1 - \frac{J'}{J}(1-\gamma' _{\bf q})-\frac{\omega}{2J}
\end{array} \right ],
\end{equation} 
where $\gamma_{\bf q} = (\cos q_x + \cos q_y)/2$ and 
$\gamma' _{\bf q} = \cos q_x \cos q_y$.
Here $J=4t^2/U$ and $J'=4t^{'2}/U$ are the NN and NNN
spin couplings in the equivalent Heisenberg model,
and the NNN term $J'(1-\gamma' _{\bf q})$ directly leads to a softening
of the magnon mode energies.
Substituting in the RPA expression, we finally obtain for the
transverse spin propagator
\begin{equation}
[\chi^{-+}({\bf q}\omega)] =
-\frac{1}{2}
\left (\frac{2J}{\omega_{\bf q}} \right ) \left [
\begin{array}{cc}
1 - \frac{J'}{J}(1-\gamma ' _{\bf q})-\frac{\omega}{2J} & -\gamma_{\bf q} \\
-\gamma_{\bf q} & 1 - \frac{J'}{J}(1-\gamma' _{\bf q}) + \frac{\omega}{2J}
\end{array} \right ]
. \left ( \frac{1}{\omega-\omega_{\bf q} + i \eta}
- \frac{1}{\omega + \omega_{\bf q} -i\eta}
\right ),
\end{equation} 

\begin{multicols}{2}\narrowtext
\noindent
where the magnon-mode energy $\omega_{\bf q}$ is given by
\begin{equation}
\left ( \frac{\omega_{\bf q}}{2J} \right )^2=
\left \{ 1-\frac{J'}{J}(1-\gamma' _{\bf q}) \right \}^2
- \gamma_{\bf q} ^2.
\end{equation}

In the long wavelength limit $(q\rightarrow 0)$,
with $\gamma' _{\bf q} = \cos q_x \cos q_y  \approx 
(1-q_x ^2 /2)(1-q_y ^2 /2) = 1-q^2/2$,
and $\gamma_{\bf q} = (\cos q_x + \cos q_y)/2 \approx 1-q^2/4$,
the magnon energy reduces to:
\begin{equation}
\omega_{\bf q}=\sqrt{2} J q 
\left ( 1-\frac{2J'}{J} \right )^{1/2}
\end{equation}
showing the strong softening of low-energy modes
by the NNN hopping. The spin-wave velocity vanishes in the
limit $J'/J \rightarrow 1/2$.
The magnon density of states 

\begin{figure}
\vspace*{-60mm}
\hspace*{-28mm}
\psfig{file=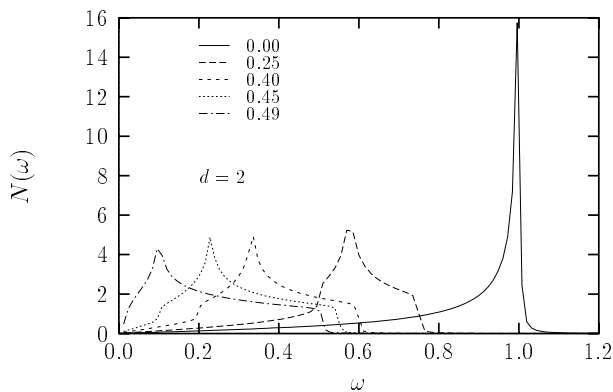,width=135mm,angle=0}
\vspace{-70mm}
\caption{
The magnon density of states for different values of the
ratio $J'/J$.}
\end{figure}

\noindent
evaluated from Eq. (15) is shown in Fig. 8 for different 
values of the ratio
$J'/J$. NNN hopping clearly softens
the magnon energies,
and transfers the magnon spectral
weight from the high-energy to the low-energy region.

The strong softening of the magnon-mode energies
suggests an enhancement in the transverse spin fluctuations.
To examine this we evaluate the transverse spin correlations
as described earlier.\cite{trans,neel}
From Eq. (14) for the transverse spin propagator,
after performing the frequency integral we obtain
the local transverse spin correlations
\begin{equation}
\langle S^+ S^- + S^- S^+ \rangle_{\rm RPA} =\sum_{\bf q}
\frac{2J}{\omega_{\bf q}}
\left \{ 1-\frac{J'}{J}(1-\gamma' _{\bf q})\right \} .
\end{equation}
The spin-fluctuation correction to sublattice magnetization is
then obtained from\cite{trans,neel}
\begin{equation}
\delta m_{\rm SF} =
\frac{\langle S^+ S^- + S^- S^+ \rangle_{\rm RPA} }
{\langle S^+ S^- - S^- S^+ \rangle_{\rm RPA} }
-1
\end{equation}
where the denominator 
$\langle S^+ S^- - S^- S^+ \rangle_{\rm RPA}$ is precisely
1 for the $S=1/2$ system due to the commutation
relation $[S^+,S^-]=2S^z$.\cite{trans,neel}
The spin-fluctuation correction to sublattice magnetization,
evaluated from Eqs. (17), (18) is shown in Fig. 9,
showing the rapid rise in transverse spin fluctuations
with the frustrating NNN spin coupling $J'$.

\begin{figure}
\vspace*{-60mm}
\hspace*{-28mm}
\psfig{file=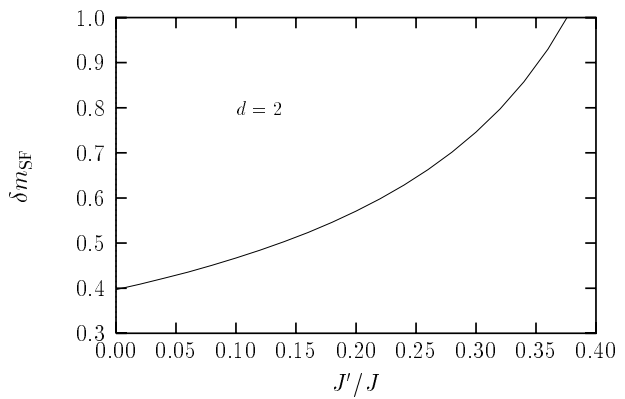,width=135mm,angle=0}
\vspace{-70mm}
\caption{
The rapid increase in the spin-fluctuation correction to sublattice
magnetization with the frustrating NNN spin coupling $J'$.}
\end{figure}

\begin{figure}
\vspace*{-60mm}
\hspace*{-28mm}
\psfig{file=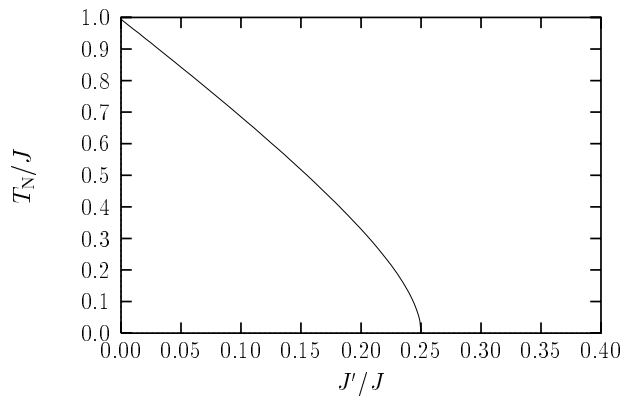,width=135mm,angle=0}
\vspace{-70mm}
\caption{
The rapid decrease in the N\'{e}el temperature for the
simple cubic lattice with the frustrating NNN spin coupling $J'$.
$T_{\rm N}/J=0.989 $ for $J'=0$.}
\end{figure}

We now consider the reduction in the N\'{e}el temperature
in three dimensions due to the frustrating NNN spin coupling.
For this purpose we consider the NNN hopping model on a simple cubic
lattice. In this case the lattice free-particle energies are
\begin{eqnarray}
\epsilon_{\bf k} &=& -2t(\cos k_x +\cos k_y +\cos k_z), \nonumber \\
\epsilon '_{\bf k} &=& -4t'(\cos k_x \cos k_y + \cos k_y \cos k_z +
\cos k_z \cos k_x) \; ,
\end{eqnarray}
and simple extension of the earlier treatment for the two-dimensional case 
leads to the following result for the
transverse spin propagator at the RPA level 
\end{multicols}

\widetext
\begin{equation}
[\chi^{-+}({\bf q}\omega)] =
-\frac{1}{2}
\left (\frac{3J}{\omega_{\bf q}} \right ) \left [
\begin{array}{cc}
1 - \frac{2J'}{J}(1-\gamma ' _{\bf q})-\frac{\omega}{3J} & -\gamma_{\bf q} \\
-\gamma_{\bf q} & 1 - \frac{2J'}{J}(1-\gamma' _{\bf q}) + \frac{\omega}{3J}
\end{array} \right ]
. \left ( \frac{1}{\omega-\omega_{\bf q} + i \eta}
- \frac{1}{\omega + \omega_{\bf q} -i\eta}
\right ),
\end{equation} 
\begin{multicols}{2}\narrowtext
\noindent
where
$\gamma_{\bf q}=(\cos q_x +\cos q_y +\cos q_z)/3$ and 
$\gamma' _{\bf q}=(\cos q_x \cos q_y + \cos q_y \cos q_z +
\cos q_z \cos q_x)/3$.
The magnon-mode energy $\omega_{\bf q}$ is given by
\begin{equation}
\omega_{\bf q}=3J \left [
\left \{1-\frac{2J'}{J}(1-\gamma' _{\bf q})\right \}^2 -\gamma_{\bf q} ^2
\right ]^{1/2}  .
\end{equation}
For small $q$,
with $\gamma' _{\bf q}  \approx 1-q^2/3$
and $\gamma_{\bf q} \approx 1-q^2/6$,
the magnon energy reduces to
\begin{equation}
\omega_{\bf q}=\sqrt{3} J q 
\left (1-\frac{4J'}{J} \right )^{1/2}
\end{equation}
which vanishes in the limit $J'/J \rightarrow 1/4$
due to the frustration effect of the NNN coupling $J'$.
The softening of the low-energy magnon spectrum
has a bearing on the N\'{e}el temperature,
as discussed below.

Within the renormalized spin-fluctuation theory,\cite{neel}
the N\'{e}el temperature $T_{\rm N}$
is obtained from the isotropy condition
$\langle S^+ S^- + S^- S^+ \rangle_{T=T_{\rm N}} = \frac{2}{3} S(S+1) $.
For the NN coupling model ($J'=0$),
the N\'{e}el temperature was obtained earlier
as $T_{\rm N}=zJ\frac{S(S+1)}{3}f_{\rm SF} ^{-1}$
for the general case of spin $S$ and $z$ nearest neighbors
on a hypercubic lattice.\cite{neel}
For the simple cubic lattice the spin-fluctuation factor 
$f_{\rm SF} \equiv \sum_{\bf q} 1/(1-\gamma_{\bf q} ^2)
=1.517 $, and for $S=1/2$ this reduces to
$T_{\rm N}/J=0.989 $. Extending this analysis to the present case,
from the expression for the magnon propagator in Eq. (20) we obtain
\begin{equation}
T_{\rm N}=
\frac{3J}{2}
\left [
\sum_{\bf q}
\frac{1-\frac{2J'}{J}(1-\gamma '_{\bf q} )}
{\left \{1-\frac{2J'}{J}(1-\gamma '_{\bf q})\right \}^2 -\gamma_{\bf q} ^2 }
\right ]^{-1}
\end{equation}

The N\'{e}el temperature, evaluated from the above equation,
is shown in Fig. 10 as a function of $J'$.
The rapid reduction of $T_{\rm N}$ with $J'$
and the vanishing at $J'=J/4$
is due to the enhancement of transverse spin fluctuations
arising from the frustration-induced softening of the long-wavelength,
low-energy magnon modes. 
The instability at $J' = J/4$ is towards a F-AF phase
with ordering wavevector ${\bf Q}=(\pi,\pi,0)$,
involving antiferromagnetic ordering 
of spins in planes (say parallel to the x-y plane),
and ferromagnetic alignment along the z direction.\cite{tbp}

\end{multicols}
\end{document}